\documentclass[twocolumn,showpacs,showkeys,reprint,amsmath,amssymb,aps]{revtex4-1}

\usepackage{graphicx}
\usepackage{dcolumn}
\usepackage{bm}
\usepackage{tabularx}
\usepackage{amsmath}

\begin{document}

\title{Thermal conductivity of triangular-lattice antiferromagnet Na$_2$BaCo(PO$_4$)$_2$: Absence of itinerant fermionic excitations}

\author{Y. Y. Huang,$^1$ D. Z. Dai,$^1$ C. C. Zhao,$^1$ J. M. Ni,$^1$ L. S. Wang,$^1$ B. L. Pan,$^1$ B. Gao,$^2$ Pengcheng Dai,$^2$ and S. Y. Li$^{1,3,4*}$}

\affiliation
{$^1$State Key Laboratory of Surface Physics, and Department of Physics, Fudan University, Shanghai 200438, China\\
 $^2$Department of Physics and Astronomy, Rice University, Houston, Texas 77005, USA\\
 $^3$Collaborative Innovation Center of Advanced Microstructures, Nanjing 210093, China\\
 $^4$Shanghai Research Center for Quantum Sciences, Shanghai 201315, China
}

\date{\today}

\begin{abstract}
We present the ultralow-temperature specific heat and thermal conductivity measurements on single crystals of triangular-lattice antiferromagnet Na$_2$BaCo(PO$_4$)$_2$, which was recently argued to host itinerant fermionic excitations, like a quantum spin liquid, above its antiferromagnetic phase transition temperature $T_{\rm N}$ = 0.148 K. In specific heat measurements, we confirm the peaks due to antiferromagnetic ordering when magnetic field $\mu_0 H \leq$ 1 T, roughly consistent with previous work [N. Li $et$ $al.$, Nat. Commun. 11, 4216 (2020)]. However, in thermal conductivity measurements, we observe negligible residual linear term in zero and finite magnetic fields, in sharp contrast to previous report [N. Li $et$ $al.$, Nat. Commun. 11, 4216 (2020)]. At 0.35 K, the thermal conductivity increases with field up to 3 T then saturates, similar to that of another triangular-lattice compound YbMgGaO$_4$, which further shows that the heat is conducted only by phonons with scattering from spins and boundary. Our results clearly demonstrate the absence of itinerant fermionic excitations in the disordered state above $T_{\rm N}$ in this frustrated antiferromagnet Na$_2$BaCo(PO$_4$)$_2$, thus such a state is not as exotic as previously reported.
\end{abstract}

\maketitle

Motivated by Anderson's pioneering proposal \cite{Anderson1973,Anderson1987}, people have long sought for quantum spin liquid (QSL) state experimentally \cite{BalentsQSL,ZhouQSL,SavaryQSL,BroholmQSL}. QSL state is an exotic ground state with strong quantum fluctuations preventing spins from ordering and thus keeping them in a liquid-like state even at absolute zero termperature \cite{BalentsQSL}. Fractionalized excitations, such as spinons, are the most important characteristic in a QSL state and detecting them will give crucial information on this exotic state \cite{ZhouQSL,SavaryQSL}. Geometrical frustration like triangular and kagome lattice can easily provide fertile soil for the quantum fluctuations in QSL state and the spin-1/2 triangular-lattice compounds have been most studied because of its simplest frustrated structure \cite{Anderson1973,BalentsQSL,ZhouQSL}. Among those triangular-lattice QSL candidates, the two organic compounds $\kappa$-(BEDT-TTF)$_2$Cu$_2$(CN)$_3$ and EtMe$_3$Sb[Pd(dmit)$_2$]$_2$, and the inorganic compound YbMgGaO$_4$ are some typical examples, although their exact ground states are still under debate \cite{saltfirst,saltHC,saltkappa,dmitfirst,dmitkappa,dmitnmr,dmitHC,dmitNi,dmitlouis,YMGOfirst,YMGOneutron1,YMGOneutron2,YMGOkappa,YMGOglass}.

Longitudinal thermal conductivity measurements have been proved to be a direct technique to detect elementary excitations in QSL or QSL-like candidates \cite{saltkappa,dmitkappa,dmitNi,dmitlouis,YMGOkappa}. A finite residual linear term $\kappa_0/T$ is a smoking-gun evidence for the existence of itinerant gapless spinons \cite{theoryLee,theoryNave,theoryZhou,theoryWerman}. However, $\kappa_0/T$ of above QSL candidates is either negligible \cite{saltkappa,YMGOkappa}, or highly controversial \cite{dmitkappa,dmitNi,dmitlouis}.

Recently, Na$_2$BaCo(PO$_4$)$_2$, a new triangular-lattice antiferromagnet with spin-1/2 Co$^{2+}$ ions was proposed and synthesized successfully \cite{NBCPOcava}. The magnetic Co$^{2+}$ ions and nonmagnetic Ba$^{2+}$ ions in Na$_2$BaCo(PO$_4$)$_2$ form alternating triangular planes along the crystallographic $c$ axis in a simple A-A-A stacking pattern, as shown in Fig. 1(a). Disorder due to the site mixing is avoided thanks to the distinct ionic size and much weaker exchange coupling between interlayer Co$^{2+}$ ions guarantees the perfect two-dimensional triangular structure of Co$^{2+}$ ions, as depicted in Fig. 1(b). No magnetic ordering above 0.3 K was found in magnetic susceptibility, specific heat, and neutron scattering experiments, while localized low-energy spin fluctuations were revealed at low temperatures \cite{NBCPOcava}.

Soon after the discovery, the specific heat down to 0.05 K measured by N. Li $et$ $al.$ clearly showed that Na$_2$BaCo(PO$_4$)$_2$ undergoes antiferromagnetic transition at $T_{\rm N}$ = 0.148 K \cite{NBCPOsun}. Further susceptibility measurements as well as thermal conductivity measurements demonstrated a series of four spin state phases represented by an up-up-down phase \cite{NBCPOsun}. Moreover, a finite residual linear term $\kappa_0/T$ of 62 $\mu$W K$^{-2}$ cm$^{-1}$ was strikingly observed by extrapolating the thermal conductivity data above $T_{\rm N}$ in zero field, which vanishes at 14 T due to the spin polarization \cite{NBCPOsun}. As mentioned above, this result is a strong evidence for the existence of itinerant gapless spinons \cite{theoryLee,theoryNave,theoryZhou,theoryWerman}, making Na$_2$BaCo(PO$_4$)$_2$ different from the rest of QSL and QSL-like candidates. Based on this work, theoretical calculations revealed that the anisotropic ferromagnetic Kitaev exchange plays an important role in assisting the realization of this QSL-like state above $T_{\rm N}$ \cite{NBCPOESR}. Since the issue of detecting spinons is the foundation for understanding the nature of QSL states, it is necessary to revisit these thermodynamic and heat transport properties of Na$_2$BaCo(PO$_4$)$_2$.

\begin{figure}
\includegraphics[clip,width=7.5cm]{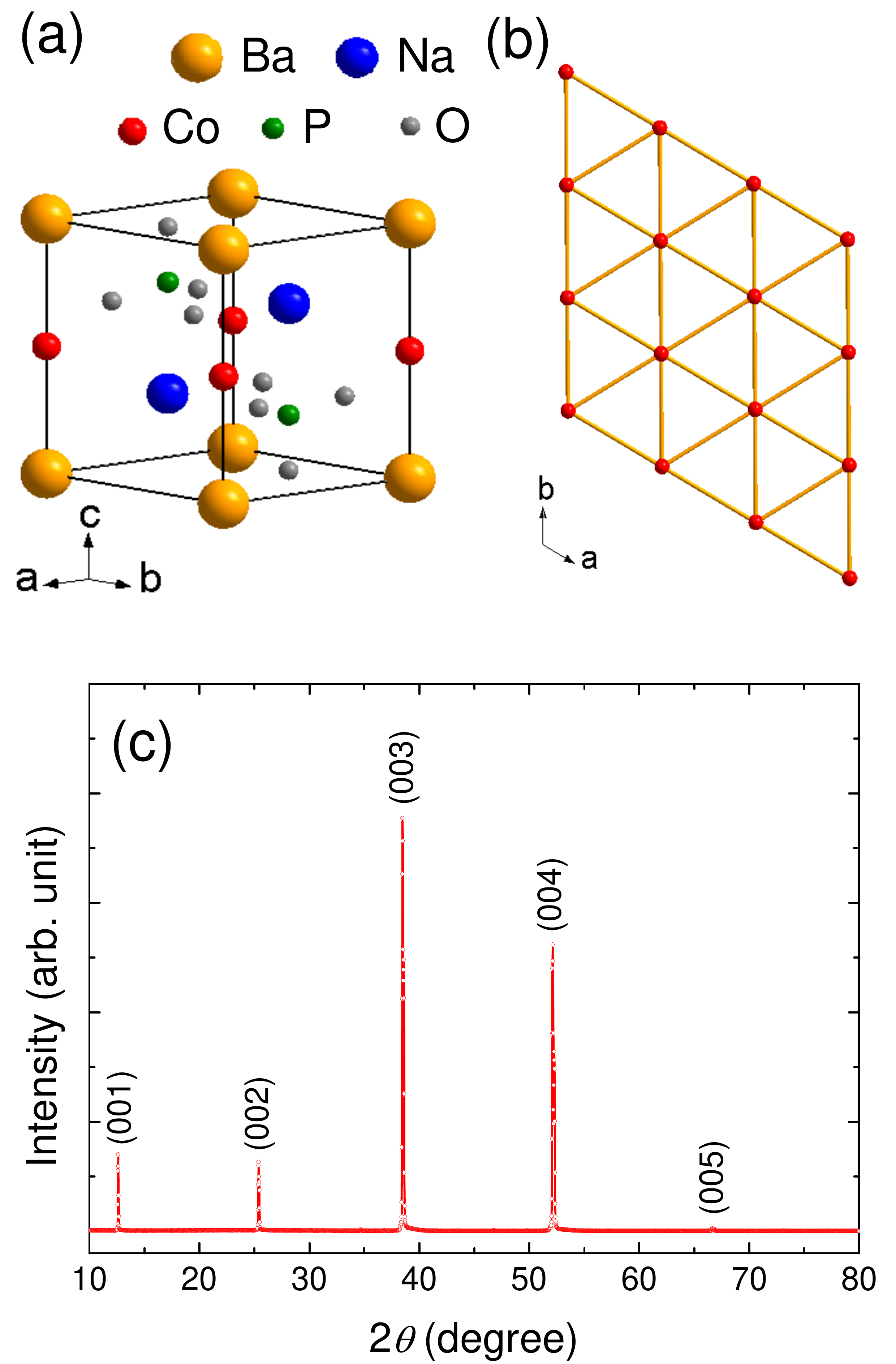}
\caption{(a) Crystal structure of Na$_2$BaCo(PO$_4$)$_2$. The Na, Ba, Co, P and O atoms are presented as blue, yellow, red, green and grey balls, respectively. (b) Triangular structure of the Co$^{2+}$ ions layer in the $ab$ plane. (c) Typical room-temperature XRD pattern from the large natural surfaces of the Na$_2$BaCo(PO$_4$)$_2$ single crystals. Only (00$l$) Bragg peaks are found.}
\end{figure}

\begin{figure}
\includegraphics[clip,width=6.45cm]{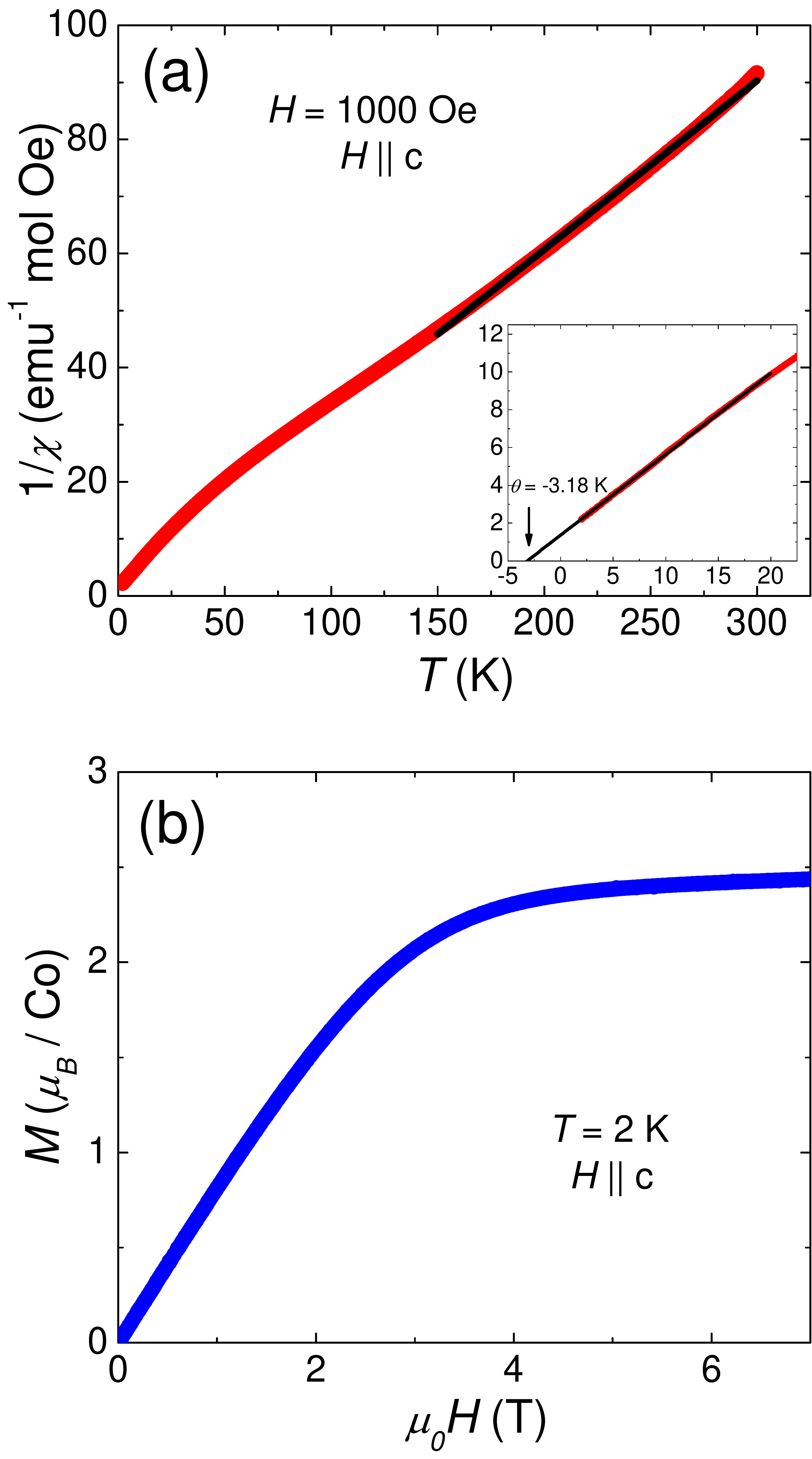}
\caption{(a) Temperature dependence of inverse magnetic susceptibility at $H$ = 1000 Oe along $c$ axis. The solid line is a Curie-Weiss fit above 150 K. Inset: zoomed-in low-temperature region. The solid line is again a Curie-Weiss fit below 20 K, giving the Curie-Weiss temperature $\theta$ = $-$3.18 K. (b) Field dependence of magnetization at $T$ = 2 K.}
\end{figure}

In this Letter, we report the ultralow-temperature specific heat and thermal conductivity measurements on high-quality single crystals of Na$_2$BaCo(PO$_4$)$_2$. In our specific heat measurements, the peak due to antiferromagnetic ordering is observed at $T_{\rm N}$ = 139 mK in zero field and does not disappear until a magnetic field of at least 1 T is applied. While our specific heat results are similar to previously reported ones, we cannot reproduce the thermal conductivity results in the same work \cite{NBCPOsun}. In our thermal conductivity measurements, negligible residual linear terms are confirmed by two samples, implying the absence of itinerant gapless fermionic excitations. The thermal conductivity increases for $\mu_0 H \leq$ 3 T and saturates above 3 T where spins are almost fully polarized. The antiferromagnetic ordering has little effect on the behaviour of thermal conductivity, which is only discernable under 0.5 T when the specific heat peak is the highest. In light of our present work, the state above $T_{\rm N}$ and its excitations should be reconsidered.

High-quality plate-like Na$_2$BaCo(PO$_4$)$_2$ single crystals were grown from NaCl-flux as previously described \cite{NBCPOcava}. The X-ray diffraction (XRD) measurements were performed by an X-ray diffractometer (D8 Advance, Bruker). The largest natural surface of the as-grown single crystals was determined as (001) plane by XRD, as illustrated in Fig. 1(c). The magnetic susceptibility measurement was performed using a magnetic property measurement system (MPMS, Quantum design) and the specific heat was measured down to 90 mK via the relaxation method in a physical property measurement system (PPMS, Quantum Design) equipped with a small dilution refrigerator. Two samples, labeled as 1 and 2, with dimensions of 2.00 $\times$ 1.25 $\times$ 0.18 mm$^3$ and 2.14 $\times$ 0.62 $\times$ 0.18 mm$^3$ were used for the heat transport measurements. The in-plane thermal conductivity was measured in a large dilution refrigerator by using a standard four-wire steady-state method with two RuO$_2$ chip thermometers, calibrated $in$ $situ$ against a reference RuO$_2$ thermometer. Magnetic fields were applied perpendicular to the $ab$ plane in all measurements.

Figure 2(a) shows the inverse magnetic susceptibility ($1/\chi$) from 2 K to 300 K at $H$ = 1000 Oe along $c$ axis. The linear Curie-Weiss fit above 150 K gives the effective moment of a Co$^{2+}$ ion at high temperature being 5.20 $\mu_{B}$. The deviation from the linear fit at high temperature around 50 K indicates a crossover to Kramers doublets with an effective spin $S$ = 1/2 for Co$^{2+}$ ions. Another Curie-Weiss fit is made from 2 K to 20 K in the inset of Fig. 2(a), giving the effective moment of a Co$^{2+}$ ion at low temperature being 4.32 $\mu_{B}$ and the intrinsic spin-1/2 exchange Curie-Weiss temperature $\theta$ = $-$3.18 K. The $g$ factor of 4.99 deduced thereout is close to the magnetization and electron spin resonance results in Ref. \cite{NBCPOESR}. Assuming Na$_2$BaCo(PO$_4$)$_2$ satisfies the XXZ model like YbMgGaO$_4$ \cite{intercalculate}, we obtain the exchange interaction $J_{zz}$ = $-2\theta/3$ = 2.12 K. Field dependence of magnetization at $T$ = 2 K is plotted in Fig. 2(b). The magnetization saturates around 4 T, indicating the spins have been fully polarized at $\mu_0 H \geq$ 4 T.

Temperature dependence of the specific heat from 0.09 K to 50 K in zero field and to 30 K in various magnetic fields is displayed in Fig. 3(a). The data are plotted again in a logarithmic scale in the inset of Fig. 3(a) for clarity in the low-temperature region. A small peak due to antiferromagnetic ordering is found at 139 mK in zero field. The highest peak shows up under 0.5 T at 271 mK and the peak becomes lower and shifts to 252 mK under 1 T, while no peak can be observed under 2 T and above. Magnetic humps are seen in all fields and the upturns near 100 mK result from nuclear contributions. Our results are roughly consistent with the previous work, except that N. Li $et$ $al.$ found the highest peak at 1 T \cite{NBCPOsun}.

\begin{figure}
\includegraphics[clip,width=6.65cm]{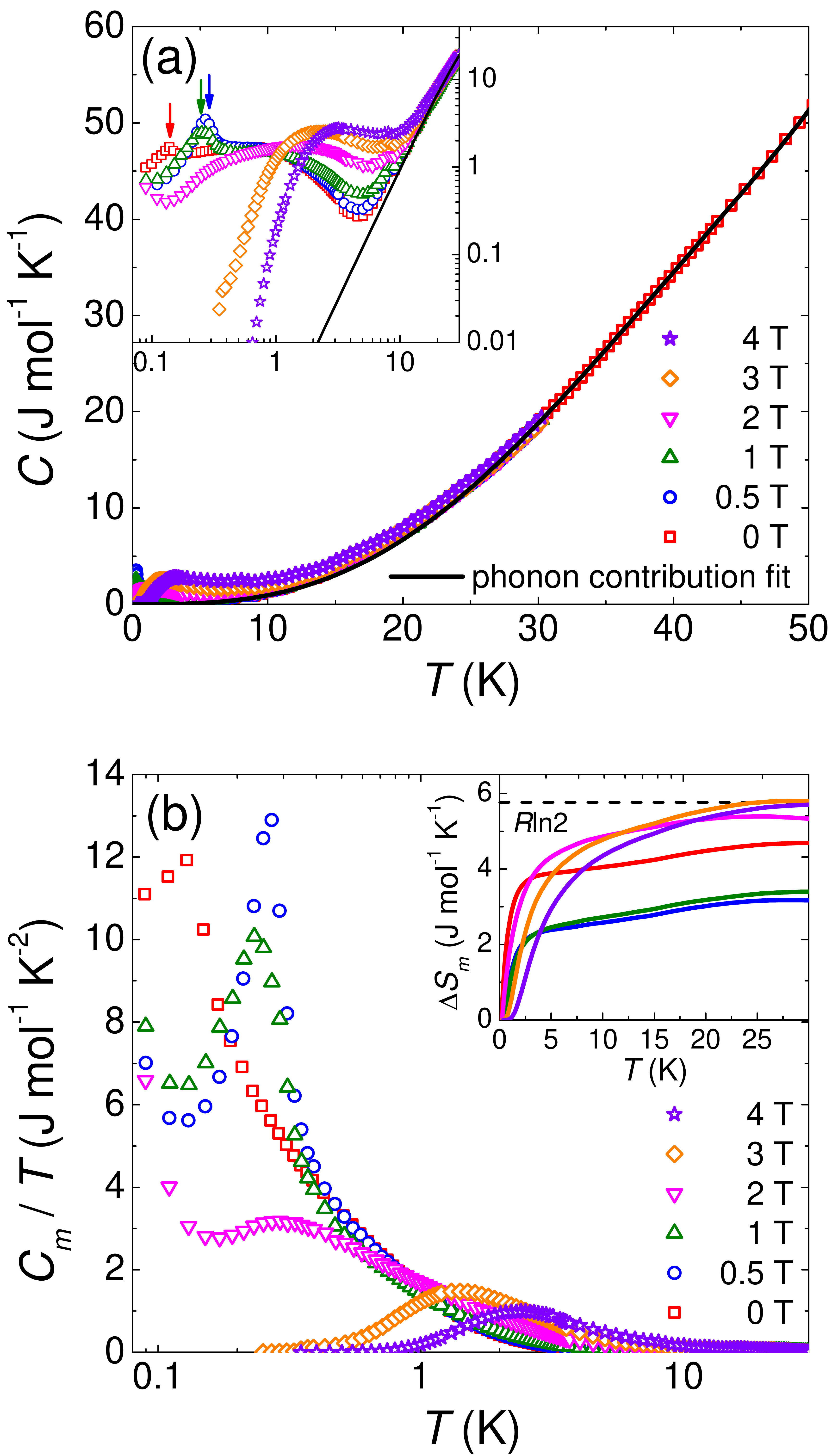}
\caption{(a) Temperature dependence of specific heat in magnetic fields up to 4 T. The solid line is a phonon contribution fit to $C_{ph}$ = $\beta T^3$ + $\gamma T^5$ + $\delta T^7$ between 30 K and 50 K in zero field. Inset: same data plotted in a logarithmic scale. The peaks attributed to the formation of an antiferromagnetic order are clearly observed at $\mu_0 H \leq$ 1 T and indicated by arrows in the same colors with the curves. The peaks locate at 139 mK, 271 mK and 252 mK under 0 T, 0.5 T and 1 T, respectively. Magnetic humps are seen in all fields and the upturns near 100 mK result from nuclear contributions. (b) Temperature dependence of magnetic specific heat divided by temperature $C_{m}/T$ under magnetic fields as above. Inset: magnetic entropies obtained by integrating the magnetic specific heat above peak or upturn temperatures. The entropies reach $R\ln2$ at 30 K under 3 T and 4 T as anticipated. }
\end{figure}

Since the data in all fields collapse to a single curve when approaching 30 K, it is reasonable to attribute the specific heat above 30 K totally to phonons. We fit the data above 30 K in zero field to $C_{ph}$ = $\beta T^3$ + $\gamma T^5$ + $\delta T^7$ and obtain $\beta$ = $9.73 \times 10^{-4}$ J mol$^{-1}$ K$^{-4}$, $\gamma$ = $-3.54 \times 10^{-7}$ J mol$^{-1}$ K$^{-6}$ and $\delta$ = $5.15 \times 10^{-11}$ J mol$^{-1}$ K$^{-8}$. The magnetic specific heat $C_{m}$ is easily obtained by subtracting the phonon contribution from the data in Fig. 3(a). Figure 3(b) plots the $C_{m}/T$ versus $T$. In order to avoid the contamination of the nuclear contributions and the antiferromagnetic ordering, we integrate the $C_{m}/T$ above peak or upturn temperatures and obtain the magnetic entropies in zero and finite magnetic fields. The entropies reach $R\ln2$ at 30 K under 3 T and 4 T as anticipated for the effective spin-1/2 system, while there are more magnetic entropies present below the peak or upturn temperatures under $\mu_0 H \leq$ 2 T.

The contribution from fermionic excitations to the specific heat of Na$_2$BaCo(PO$_4$)$_2$ is hardly verified because of the contamination of antiferromagnetic ordering and nuclear contribution, while the longitudinal thermal conductivity measurement can largely get rid of these troubles. Figure 4(a) presents the in-plane ultralow-temperature thermal conductivity of two Na$_2$BaCo(PO$_4$)$_2$ single crystals (labelled as sample 1 and sample 2) in zero field, plotted as $\kappa/T$ vs. $T$. The thermal conductivity at very low temperature usually can be fitted to $\kappa/T$ = $a$ + $bT^{\alpha-1}$, where the two terms $a$ and $bT^{\alpha-1}$ represent contributions from itinerant gapless fermionic magnetic excitations (spinons here) and phonons, respectively \cite{thermalpower1,thermalpower2}. The fitting below 0.4 K gives $(\kappa_0/T)_1$ = 6 $\pm$ 3 $\mu$W K$^{-2}$ cm$^{-1}$ and $\alpha_1$ = 2.61 $\pm$ 0.04 for sample 1 and $(\kappa_0/T)_2$ = 8 $\pm$ 3 $\mu$W K$^{-2}$ cm$^{-1}$ and $\alpha_2$ = 2.64 $\pm$ 0.06 for sample 2. One may worry about the influence of the magnetic ordering at 139 mK on the fitting. Taking sample 1 for example, we also fit the data between $T_{\rm N}$ = 139 mK and 0.4 K and obtain $\kappa_0/T$ = 3 $\pm$ 8 $\mu$W K$^{-2}$ cm$^{-1}$ and $\alpha$ = 2.59 $\pm$ 0.08. The fitting curve (not shown here) is overlapped with the fitting below 0.4 K (shown in Fig. 4(a)). In fact, there is no anomaly near $T_{\rm N}$ in thermal conductivity. Therefore, it is safe to fit the data down to the lowest temperature for getting the residual linear term. If spinons do exist in Na$_2$BaCo(PO$_4$)$_2$, according to the fitting results below 0.4 K, we estimate the mean free path $l_{s1}$ = 3.5 {\AA} and $l_{s2}$ = 4.7 {\AA} by using the formula $\kappa_0/T$ = $(\pi k_{B}^{2} l_{s})/(9\hbar ad)$ \cite{dmitkappa}, where $a$ = 5.32 {\AA} and $d$ = 7.01 {\AA} are nearest-neighbor and interlayer spin distance, respectively. The ($l_{s}$)s in our measurements are comparable to the interspin distance, which clearly excludes the existence of itinerant gapless spinons.

\begin{figure}
\includegraphics[clip,width=6.6cm]{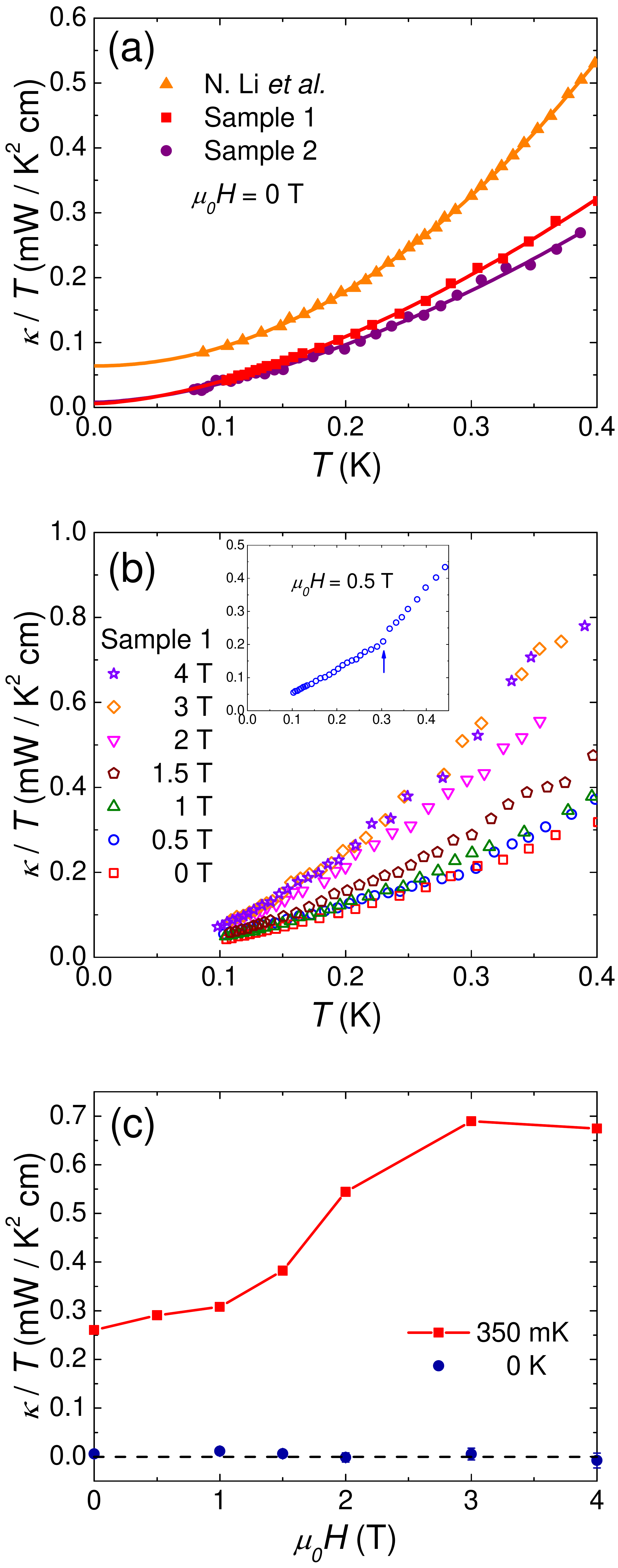}
\caption{(a) The in-plane thermal conductivity of Na$_2$BaCo(PO$_4$)$_2$ single crystals in zero field, plotted as $\kappa/T$ vs. $T$. The data in Ref. \cite{NBCPOsun} is also plotted for comparison. The solid curves represent a fit to $\kappa/T$ = $a$ + $bT^{\alpha-1}$ below 0.4 K, giving the residual linear terms $(\kappa_0/T)_1$ = 6 $\pm$ 3 $\mu$W K$^{-2}$ cm$^{-1}$ and $(\kappa_0/T)_2$ = 8 $\pm$ 3 $\mu$W K$^{-2}$ cm$^{-1}$, respectively. These values are negligible, comparing with $\kappa_0/T$ = 62 $\mu$W K$^{-2}$ cm$^{-1}$ obtained in Ref. \cite{NBCPOsun}. (b) Low-temperature in-plane thermal conductivity of sample 1 in zero and finite magnetic fields applied along the $c$ axis. Only under 0.5 T can one find a kink near the magnetic transition, as seen in the inset. The kink is indicated by an arrow at 304 mK. (c) Field dependence of residual linear term $\kappa_0/T$ and $\kappa/T$ at 350 mK, obtained by fitting the data in (b) except 0.5 T to $\kappa/T$ = $a$ + $bT^{\alpha-1}$. The $\kappa/T$ at 350 mK under 0.5 T is got by a guide to the eye.}
\end{figure}

The thermal conductivity in magnetic fields for sample 1 is shown in Fig. 4(b). Only at 0.5 T can we observe a kink at $\sim$300 mK due to the antiferromagnetic transition with the highest peak in the specific heat, as illustrated in the inset of Fig. 4(b). In order to learn the field dependence of $\kappa/T$ better, we fit the data in Fig. 4(b) except 0.5 T to $\kappa/T$ = $a$ + $bT^{\alpha-1}$ and plot the field dependence of the residual linear term $\kappa_0/T$ and $\kappa/T$ at 350 mK in Fig. 4(c). Note that the $\kappa/T$ at 350 mK under 0.5 T is got by a guide to the eye. Negligible $\kappa_0/T$ is obtained in all the fitted fields, which excludes the existence of itinerant gapless spinons in these fields. The $\kappa/T$ at 350 mK increases at low field then saturates above 3 T, reminiscent of YbMgGaO$_4$ \cite{YMGOkappa}, further demonstrates that the heat conduction in Na$_2$BaCo(PO$_4$)$_2$ is attributed to phonons scattered from spins and boundary. Generally, the fluctuating spins in frustrated magnets scatter the phonons strongly but the gradual field-induced alignment of spins can relieve this scattering effect. Indeed, the magnetization in Fig. 2(b) shows that the spins are almost fully polarized under 3 T, corresponding to the saturation of $\kappa/T$ here. Therefore, the phonons suffer less scattering from the spins and contribute more to the heat conduction when the field is increased until 3 T.

The zero-field thermal conductivity result in Ref. \cite{NBCPOsun} is also displayed in Fig. 4(a). No obvious anomaly is seen near $T_{\rm N}$ either, but the fitting below 0.4 K gives a finite $\kappa_0/T$ = 62 $\mu$W K$^{-2}$ cm$^{-1}$ and $\alpha$ $\simeq$ 3. At the first glance, one can readily find that the fitting even in the up-up-down order state alone gives almost the same value of $\kappa_0/T$. However, to our knowledge, only gapped fractionalized excitations have been found by inelastic neutron scattering (INS) in an magnetic order state of frustrated magnets like Cs$_2$CuCl$_4$ and Ba$_3$CoSb$_2$O$_9$ \cite{CsCuClins,BCSOins1,BCSOins2}, which give no residual linear term in thermal conductivity due to the existence of gap. It becomes a big question which kind of gapless fermionic excitations contribute to the finite $\kappa_0/T$ in the up-up-down order state of their Na$_2$BaCo(PO$_4$)$_2$ sample. Furthermore, the power $\alpha$ of phonon thermal conductivity is usually less than 3 at ultralow temperature. This may result the specular reflections at sample surfaces \cite{thermalpower1,thermalpower2}, or from spin-phonon scattering in some frustrated magnets. For example, $\alpha$ is even less than 2 for EtMe$_3$Sb[Pd(dmit)$_2$]$_2$, YbMgGaO$_4$ and Tb$_2$Ti$_2$O$_7$ \cite{YMGOkappa,dmitlouis,TbTiOkappa}. In this context, $\alpha$ $\simeq$ 2.6 we obtain for Na$_2$BaCo(PO$_4$)$_2$ is more reasonable than the $\alpha$ = 3 obtained in Ref. \cite{NBCPOsun}. The reason for the discrepancy between Ref. \cite{NBCPOsun} and our work is unclear to us at this stage.

Finally, we discuss the two possibilities of the true state above $T_{\rm N}$ in Na$_2$BaCo(PO$_4$)$_2$. One trivial scenario is that no well-defined gapless fermionic excitations exist in this state. The humps above $T_{\rm N}$ seen in the specific heat just come from those strongly fluctuating spins which do not contribute positively to the thermal conductivity. Another scenario is that some kind of magnetic excitations (fermionic or bosonic) do exist in this state, but they are gapped like in $\kappa$-(BEDT-TTF)$_2$Cu$_2$(CN)$_3$ \cite{saltkappa}. Indeed, INS experiments at 0.3 K on Na$_2$BaCo(PO$_4$)$_2$ do show a spin gap $\sim$ 2 K between 0 T and 2 T and larger ones at higher fields for characteristic magnetic
excitations \cite{NBCPOcava}. In this case, the humps above $T_{\rm N}$ are also well interpreted by the gapped excitations. Those gapped spin excitations will not contribute to thermal conductivity, which is consistent with our results. Nuclear magnetic resonance or more detailed INS measurements are needed to clarify the true state above $T_{\rm N}$ in Na$_2$BaCo(PO$_4$)$_2$.

In summary, we have revisited the thermodynamic and heat transport properties of triangular-lattice antiferromagnet Na$_2$BaCo(PO$_4$)$_2$. The specific heat peaks under $\mu_0 H \leq$ 1 T caused by antiferromagnetic transitions are well reproduced, as in previous work \cite{NBCPOsun}. However, no contribution from itinerant gapless fermionic excitations is observed in thermal conductivity, which contradicts the finite residual linear term reported in Ref. \cite{NBCPOsun}. The $\kappa/T$ at 0.35 K increases with magnetic field and saturates at 3 T due to the spin polarization, which further demonstrates the heat is conducted by phonons scattered by spin system and boundary. Although we exclude the existence of itinerant gapless fermionic excitations in Na$_2$BaCo(PO$_4$)$_2$, more experiments are still required to elucidate the excitations in the disordered state above $T_{\rm N}$. This is an important issue on the way pursuing the ideal QSL, when the real materials are nonideal and show antiferromagnetic order at very low temperature.

This work is supported by the Natural Science Foundation of China (Grant No: 12034004) and the Shanghai Municipal Science and Technology Major Project (Grant No. 2019SHZDZX01). Single crystal and characterization efforts at Rice are supported by the US Department of Energy, Basic Energy Sciences, under grant no. DE-SC0012311 and by the Robert A. Welch Foundation grant no. C-1839, respectively (P.D.).\\

\noindent
{$^*$ E-mail: shiyan$\_$li$@$fudan.edu.cn}

\end{document}